\documentclass[5p,preprint,authoryear]{elsarticle}

\usepackage[utf8]{inputenc}
\usepackage{hyperref}
\usepackage{amsmath}
\usepackage{bbold}

\newcommand{\abs}[1]{\ensuremath{\left|#1\right|}}
\newcommand{\quotes}[1]{``#1''}

\journal{Astronomy and Computing}

\begin{document}
	\begin{frontmatter}
		\title{SPH to Grid: a new integral conserving method}
		\author[label1]{R\"ottgers, B.}
		\ead{bernhard.roettgers@yahoo.de}
		\author[label2,label3]{Arth, A.}
		\ead{arth@usm.uni-muenchen.de (Corresponding author)}
		\address[label1]{Max-Planck-Institute for Astrophysics, Karl-Schwarzschild-Str. 1, 85741 Garching, Germany}
		\address[label2]{University Observatory Munich, Scheinerstraße 1, 81679 Munich, Germany}
		\address[label3]{Max-Planck-Institute for Extraterrestrial Physics, Giessenbachstrasse 1, 85748 Garching, Germany}
		
		\begin{abstract}
			Analysing data from Smoothed Particle Hydrodynamics (SPH) simulations is about understanding global fluid properties rather than individual fluid elements. Therefore, in order to properly understand the outcome of such simulations it is crucial to transition from a particle to a grid based picture. In this paper we briefly summarise different methods of calculating a representative volume discretisation from SPH data and propose an improved version of commonly used techniques. We present a possibility to generate accurate 2D data directly without the CPU time and memory consuming detour over a 3D grid. We lay out the importance of an accurate algorithm to conserve integral fluid properties and to properly treat small scale structures using a typical galaxy simulation snapshot. For demonstration purposes we additionally calculate velocity power spectra and as expected find the main differences on small scales. Finally we propose two new multi-purpose analysis packages which utilise the new algorithms: Pygad and SPHMapper.
		\end{abstract}
		
	\end{frontmatter}
	
	\section{Introduction}
		\label{sec:intro}
		
		Since several decades Smoothed Particle Hydrodynamics (SPH), originally formulated by \cite{Lucy1977} and \cite{Gingold1977}, is a numerical technique particularly widely used in astrophysics as well as for example engineering, geophysics and computer graphics. SPH discretises a medium and also the related equations of hydrodynamics (and potential extensions such as magnetohydrodynamics), formally derived from the fluid Lagrangian by mass \citep[see e.g.][]{Rosswog2007}. For a recent review we refer to \cite{Price2012}.
		
		To the present day many SPH codes have been written and several modifications and improvements have turned up to solve the intrinsic problems that SPH has by construction. These codes include for example Gasoline \citep{Wadsley2004}, Gadget \citep{Springel2005a}, Magma \citep{Rosswog2007}, Vine \citep{Wetzstein2009}, Phantom \citep{Price2010,Lodato2010} and Gandalf \citep{Hubber2017}. Much work has been put into improvements of the formalism targeting convergence, fluid mixing and related issues. See for example \cite{Hu2014,Beck2016}. While these cover mainly improvements of the simulation codes themselves, amongst others \cite{Beck2016} hint to the importance of a proper SPH to grid transformation scheme, since SPH shows only what happens to one fluid parcel while gridded data shows what happens to a domain. Due to the inherent particle noise it is advisable to carry out some post-processing even at the location of the original SPH particles \citep{Springel2010}. Therefore, this paper targets the question of how to properly understand and analyse the resulting SPH data.
		
		The amount of analysis codes is at least as vast as the number of simulation codes itself. To understand a SPH dataset properly and to translate the fluid variables into a commonly understandable form is far from trivial. Since SPH particles discretise mass and not volume, one can not simply plot, for instance, a column-density map of an astrophysical SPH simulation like of data produced by a grid code. While it is straight forward to think in terms of particles because it resembles a micro-physical point of view, one may never forget that SPH particles are not simply elementary particles of the fluid but rather artificial elements tracing the fluid properties like density or temperature. To visualise fluid properties in a volume related way or to perform further calculations one often needs to generate a scalar or vector field describing the respective quantity as a function of the position at first. Although the applied methods are usually not mentioned in most scientific articles, it is extremely important to conserve as much accuracy as possible within a reasonable amount of computing time since any error made in this process propagates into further analysis. Recent articles about map making and especially visualisation include for example \cite{Dolag2005,Navratil2007,Fraedrich2010,Forgan2010,Koepferl2016}.
		
		In this paper, after revisiting required SPH basics in section \ref{sec:sphbasics}, we review a few of the commonly used techniques how to transform SPH data to gridded data, display issues which arise in certain cases and discuss possibilities to solve these issues (section \ref{sec:differentmethods}). We present a novel formulation we call S-normed SPH binning, hereafter {\bf SNSB}, which approaches typical problems in a clever way while maintaining the resolution of the SPH data as good as possible (section \ref{sec:snsb}). Furthermore, we show a way to improve the run time requirements of such an algorithm by an intrinsic reduction of dimensions (section \ref{sec:2d}). We compare the different methods using some astrophysical simulation results in section \ref{sec:comparison} and conclude with a short presentation of two novel multi-purpose analysis tools which incorporate these methods and many more convenient features in section \ref{sec:tools}.

	\section{SPH principles}
		\label{sec:sphbasics}
		
		We review some SPH basics required. For an extensive review of SPH we refer to the review of \cite{Price2012}. We start with the definition of the (symmetric) smoothing kernel following \cite{Dehnen2012}:
		
		\begin{align}
			W\big(\vec r, h\big) = H (h)^{-\nu} \, w \left( \frac{|\vec r|}{H(h)} \right)
			\label{eq:kernel_def}
		\end{align}
		
		with $h$ the smoothing length (a measure for the degree of smoothing, typically direct proportional to $H$), $H$ the compact kernel support and $\nu$ the dimensionality. The smoothing length and the kernel support radius may be, but are not necessarily the same quantity. Therefore, we explicitly distinguish the two. In our tests in section \ref{sec:comparison} we use rather low resolution data with a cubic spline kernel and 32 neighbours to save computing time. We can define the kernel approximation of a physical scalar (or similar vector) field $A(\vec r)$, i.e.\ a smoothed version:
		\begin{align}
			A(\vec r) = \int \! \mathrm d^3 \vec r\, ' \, A(\vec r\, ') \, \delta(\vec r - \vec r\, ')
			\approx \int \! \mathrm d^3 \vec r\, ' \, A(\vec r\, ') \, W(\vec r - \vec r\, ', h) \label{eq:Kernel_integ}
		\end{align}
		This integral is discretised in SPH, historically in terms of mass:
		\begin{align}
			A(\vec r) &= \int \! \underbrace{\rho(\vec r\, ') \mathrm d^3 \vec r\, '}_{\mathrm dm} \, \frac{1}{\rho(\vec r\, ')} \, A(\vec r\, ') \, W(\vec r - \vec r\, ', h) \\
			&\approx \sum_{j=1}^{N} \frac{m_j}{\rho_j} \, A_j \, W(\vec r - \vec r_j, h)  \label{eq:SPH_est_m_rho} \\
			&= \boxed{ \sum_{j=1}^{N} \Delta V_j \, A_j \, W(\vec r - \vec r_j, h) =: \langle A(\vec r)\rangle_\text{SPH} }.	\label{eq:SPH_est_dV}
\end{align}
		Generally, it can be discretised with some volume $\Delta V_j$ of the particle $j$, following \cite{Hopkins2013a}.
		
		When binning an arbitrary quantity $A(\vec r)$ onto a grid, one typically seeks to calculate the mean value $A^{(\vec k)}$ at the centre of each cell $\vec k$ (here in the scatter approach):
		\begin{align}
		A^{(\vec k)} &= \frac{1}{\Delta V^{(\vec k)}} \int\limits_{\Delta V^{(\vec k)}} \! \mathrm d^3 \vec r \, A(\vec r) \\
		&\stackrel{(\ref{eq:SPH_est_dV})}= \frac{1}{\Delta V^{(\vec k)}} \sum_{j=1}^{N} \Delta V_j \, A_j \, \int\limits_{\Delta V^{(\vec k)}} \! \mathrm d^3 \vec r \,\, W(\vec r - \vec r_j, h_j)
		\label{eq:ideal_grid_value}
		\end{align}
		with $\Delta V^{(\vec k)}$ being the volume of cell $\vec k$.
		
	\section{SPH to Grid: Different Methods}
		\label{sec:differentmethods}
	
		In this section we present some of the commonly used methods for a transition from SPH to gridded data. In order to judge on the quality of these approaches we briefly devote ourselves to the question which requirements they should fulfil in order to reproduce the same meaning as the initial data and to keep the required computation time low:
		
		\begin{itemize}
			\item Conserving integral properties between SPH and grid data.
			\item (Which implies:) Taking all particles inside the defined region into account.
			\item Maintaining the provided resolution of the given SPH data.
			\item Properly treating boundaries.
			\item Maintaining balance between a computationally cheap algorithm for post-processing and small resulting errors.
		\end{itemize}
		
		We already mentioned that SPH particles should not be taken as real physically existing particles but as moving fluid elements. Therefore, a transformation involving the actual SPH smoothing kernel most likely gives the best results in contrast to an algorithm which treats the particles directly. In the following subsections we outline several algorithms from a direct particle picture to a smoothed approach.
		
		\subsection{Particle-Mesh and Window Functions}
		
			\cite{Eastwood1974} and \cite{Cui2008} give a comprehensive overview over several so called window functions. The idea behind this class of algorithms is, to select particles in a certain range from a grid cell's centre and add their contribution to the cell's value multiplied by a certain weight. The main requirements which these are constructed to fulfil are a compact top-hat support in Fourier space and a generic compact support also in real space. While the former helps to minimise sampling effects \citep[i.e. shot noise, for further reference we refer to][]{Jing2005}, the latter is useful to restrain computational cost by keeping the number of contributing particles bound.
			
			Technically not a window function but similar nevertheless, \cite{Bauer2012} display the simplest approach by always applying the value of the closest particle to a cell without any summation. This requires the use of a grid resolution about half the minimum distance of SPH particles in order to capture all particles and leads to aliasing.
			
			For a generic window function $W$ one can calculate the value of a cell $\vec k$ via the sum over all particles $i$ as
			
			\begin{equation}
				A^{(\vec k)} = \frac{1}{\mathrm{Norm}} \cdot \sum_i A_i W(x_i) W(y_i) W(z_i)
			\end{equation}
			with $A^{(\vec k)}$ being the value of the grid cell, $A_i$ the particle's value of property $A$ and $x_i, y_i, z_i$ the particle's coordinates with respect to the cell's centre. Window functions are defined symmetrically along all three coordinate axis and the function $W$ itself is the same for each coordinate direction.
			
			Frequently used window functions are for example the {\bf N}earest {\bf G}rid {\bf P}oint method which takes all particles inside a cell into account without weighting them \citep{Hockney1966}:
			
			\begin{equation}
				W(x_i) = \begin{cases}
					1 & \abs{x_i} < 0.5 \cdot d\\
					0 & \mathrm{otherwise,}
				\end{cases}
			\end{equation}
			the {\bf T}riangular {\bf S}haped {\bf C}loud method which takes particles in range of half the neighbouring cells into account using a linear distance weighting \citep{Birdsall1969,Birdsall1970}:
			
			\begin{equation}
			W(x_i) = \begin{cases}
			1-\abs{x_i} & \abs{x_i} < 1.0 \cdot d\\
			0 & \mathrm{otherwise}
			\end{cases}
			\end{equation}
			and finally the {\bf C}loud {\bf I}n {\bf C}ell method, which further extends the range and provides a smoother second order weighting, by extending the window function \citep{Buneman1973,Hockney1981}:
			
			\begin{equation}
			W(x_i) = \begin{cases}
			0.75-x_i^2 & \abs{x_i} < 0.5 \cdot d\\
			\frac{\left(1.5 - \abs{x_i} \right)^2}{2} & 0.5 < \abs{x_i} < 1.5 \cdot d\\
			0 & \mathrm{otherwise,}
			\end{cases}
			\end{equation}
			with $d$ denoting the grid cells side length. Furthermore, we refer the interested reader to {\bf P}article {\bf I}n {\bf C}ell \citep{Buneman1959,Dawson1960,Dawson1962,Morse1969}, extensions \citep{Hockney1981,Birdsall1991} and improvements like so called Multipole NGP \citep{Kruer1973}.

			Fulfilling the requirement of a compact top hat support in Fourier space leads to a non compact support in real space. \cite{Daubechies1992} provide a solution approach for mixture of both criteria: so called wavelets. These are base functions similar to those of the Fourier transformation, consisting of sin and cos terms, however, preserving local information better than the latter.
	
		\subsection{SPH Kernel Methods}
		
			\subsubsection{General Idea}
			
				Since SPH already comes with a kernel weighting formalism with compact support, one can straightforwardly replace the sum over particles by the typical SPH sum:
				\begin{align}
					\tilde A^{(\vec k)} = \sum_{j=1}^{N} \Delta V_j \, A_j  \, W(\vec r\,{}^{(\vec k)} - \vec r_j, h_j)  \label{eq:SPHToGridBasic}
				\end{align}
				Please note the important difference to window functions which are defined such, that each particle is assigned to at least one cell due to the window being of cubic support, which is (typically) larger than a voxel. However, SPH assumes particles to have a spherical smoothing via the kernel function which can result in particles not being accounted for. We go into more detail about that in the next subsection.
				
				Besides clear differences between window functions and a kernel based approach, the choice of the kernel is equally important especially when it comes to its Fourier properties. For further reading on this we refer for example to \cite{Dehnen2012} and \cite{Beck2016} and state that, while our models are valid for all kernels, it is important for consistency to use the same kernel in post-processing as during the SPH simulation itself. In our analysis we use the cubic spline function.
				
				In principal the SPH sum results in a physically more meaningful value, since it mirrors exactly how a simulation treats the respective equations. However, some problems, for which we discuss several solution approaches in the next subsections, can still arise.
			
			\subsubsection{Solving Common Issues}
				\label{sec:problemssolutions}
			
				Without modifications this approach formally fails all criteria outlined above except being a computationally relatively cheap scheme even though it resembles very closely how SPH works. Over the years several people evaluated improvements on the algorithm in order to fix these shortcomings as best as possible.
			
				\vspace{0.3cm}	
				{\bf Multiple evaluations}
				
				So far all mentioned prescriptions assumed that we want to calculate the value in the centre of a grid cell and, therefore, weighted a particles contribution by the distance to this centre. In order to increase accuracy one can average over several points within a voxel.\footnote{A voxel is the same as a pixel but in three dimensions.} An example is given by \cite{Pakmor2006} using 9 distinct points in the cell:
				
				\begin{align}
					A^{(\vec k)} =& \frac{1}{9} \cdot \sum_i \frac{m_i \cdot A_i}{\rho_i}  \left[ W \left( r_i - r\,{}^{(\vec k)} \right) \right. \nonumber\\
					& \left. + W \left( r_i + \frac{d}{\sqrt{3}} (\pm 1, \pm 1, \pm 1) - r\,{}^{(\vec k)} \right) \right]
				\end{align}
				
				Obviously this increases the computational cost by roughly a factor of 9 but captures way more detail in a single grid cell possibly yielding better results for large variations in one cell. It is also useful not to miss certain particles, if they actually reside inside a certain grid cell but their kernels do not overlap with the cell centre. However, the results are still resolution dependent and the same effect might be reachable by simply decreasing the grid cell size, therefore we do not further discuss this approach.
				
				\vspace{0.3cm}	
				{\bf Broadening the kernel}
				
				To prevent particles from falling through the grid, \cite{Pakmor2006} also suggests the possibility to broaden the kernel, i.e. to add a constant to the smoothing length of each particle. Setting this value to $\sqrt{3}/2$ times the cell size guarantees that each particle will contribute to at least one cell in three dimensions while the effect on particles with big smoothing lengths is on average small, since the relative change of the smoothing length is small. However, this approach artificially worsens the acquired resolution of the grid, since small features are increased in size and therefore smoothed out which is a drastic drawback whenever one is interested in the details. Furthermore, an additional error term is introduced, regarding the conservation of properties when integrating over the resulting grid.
				
				\vspace{0.3cm}
				{\bf Boundaries}
				
				A serious problem is posed by the boundaries, since here again particles may easily intersect with a cell but not any point inside it. We briefly discuss this as a special case when outlying our final approach in the next section.
				
				\vspace{0.3cm}
				{\bf Normalisation of SPH noise}
				
				SPH fields have shot noise due to the particle discretisation and the fixed form of the kernels.
				This means for instance that even if all SPH particles carry the same value for an arbitrary fluid quantity, the total field is not necessarily constant but flawed by particle noise. One can divide the end result by the unity condition $I$ as suggested by \cite{Price2007}:
				
				\begin{equation}
					I(\vec r\,{}^{(\vec k)}) = \sum_i \frac{m_i}{\rho_i} W \left( \vec r_i - \vec r\,{}^{(\vec k)} \right) \approx 1
					\label{eq:unityconditions}
				\end{equation}
				
				This re-normalisation has its biggest effect at locations where only few SPH particles sit, ergo for example at the boundaries of a fluid. Without this additional term one can expect very small values at those locations and a smooth drop off while the re-normalisation increases the resulting values. While this is certainly not always wished for, it brings the benefit of independence from the particle structure. \cite{Price2007} suggests to use this approach whenever no free surfaces are involved. An additional benefit of this modification is that it allows to generate weighted data in the particles in a very straight forward way by replacing the quantity $A_i$ in the numerator by $A_i \cdot w_i$ and introducing $w_i$ in the denominator. Typically weighting is, however, done in the volume discretised picture.

				\vspace{0.3cm}				
				{\bf Conservation of integral properties}
				
				Another conservation property we can impose is that the integral over the whole volume in particle and grid picture of some quantity should always be the same. This means for example, that total mass or energy is conserved in the conversion process. We further elaborate on this condition in the next section and formulate the condition quantitatively, which whill then call `S-normalisation'\footnote{The $S$ in the name stands for sum, as the sum of the kernel values at the grid points times their volume (a discrete integral) shall be equal to one as for the continuous integral over the kernels.}. Note that this requirement does not conflict with the normalisation of SPH noise.

				\vspace{0.3cm}				
				{\bf Resolution}
				
				Finally, we want to be able to transport the same resolution we have in SPH data to the grid. This is not an easy task with a fixed grid, since SPH has the advantage of being able to model large density contrasts and therefore inhomogeneities. Mesh refinement techniques come to mind as an option, however these are beyond the scope of this work. For a fixed grid one can start by choosing the cells' side length to a value comparable to the average smoothing length. This results in general in regions with plenty particles intersecting with a cell and such where only very few or even no particles lay. While the former leads to a resolution worse than possible, the latter is what really divides the methods for binning. In contrast to an SPH formalism, window functions will drastically fail since they rely heavily on counting of particles. This can lead to the conclusion, that only a SPH-like approach grants the possibility for resolution down to particle level.

			\subsubsection{S-Normed SPH-Binning}
				\label{sec:snsb}
				
				Having outlined several conditions, problems and solution approaches for general SPH binning, we propose a different ansatz in this section which we call the {\bf S-normed SPH binning} (from here on SNSB). The basic idea behind this method is that we calculate the kernel integral discretised to the sum $S$ over the grid and require that it is equal to one as the continuous integral over the kernel:
				
				\begin{align}
				\boxed{
					S_j = \sum_{\vec k} \Delta V^{(\vec k)} \, W(\vec r\,{}^{(\vec k)} - \vec r_j, h_j) \label{eq:diskrete_kernel_int}
				}
				\end{align}
				with the sum over all grid cell centres $\vec{k}$. In the limit of an infinitesimal small cell size, $\Delta V^{(\vec k)} \to 0$, we indeed recover the continuous integral:
				\begin{align}
					S_j \stackrel{\Delta V^{(\vec k)} \to 0}{\to} \int \! \mathrm d^3 \vec r \, W(\vec r - \vec r_j, h_j) = 1
					\label{eq:slimes}
				\end{align}

				This normalisation by $S$ implies that each particle $j$ contributes fully and with the correct weight to the grid (neglecting boundaries of the grid here).

				For $S_j \neq 0$, we define the cell value $\tilde A^{(\vec k)}$ for cell $\vec k$ by modifying eq.~\ref{eq:SPHToGridBasic}:
				\begin{align}
				\boxed{
				    \left. \tilde A^{(\vec k)} \right|_{S_j \neq 0} := \sum_{j=1}^{N} \Delta V_j \, \left( S_j \right)^{-1} \! \left. A_j \right|_{S_j \neq 0} \, W(\vec r\,{}^{(\vec k)} - \vec r_j, h_j). \label{eq:grid_value_S_not_0}
					}
				\end{align}
				Please note that the formula asymptotically goes to equation \ref{eq:SPH_est_dV} for small cells since $S_j$ becomes unity. We can easily prove that the integral for an arbitrary property $A$ is conserved using this modification
				\begin{align}
					&\left. \sum_{\vec k} \Delta V^{(\vec k)} \, \tilde A^{(\vec k)} \right|_{S_j \neq 0} \\
						&\stackrel{(\ref{eq:grid_value_S_not_0})}= \sum_{\vec k} \Delta V^{(\vec k)} \, \sum_{j=1}^{N} \Delta V_j \, \left(S_j \right)^{-1} \! \left. A_j \right|_{S_j \neq 0} \, W(\vec r\,{}^{(\vec k)} - \vec r_j, h_j) \\
						&= \sum_{j=1}^{N} \Delta V_j \, \left. A_j \right|_{S_j \neq 0} \, \left(S_j \right)^{-1} \underbrace{\sum_{\vec k} \Delta V^{(\vec k)} \, W(\vec r\,{}^{(\vec k)} - \vec r_j, h_j)}_{= S_j} \\
						&= \sum_{j=1}^{N} \Delta V_j \, \left. A_j \right|_{S_j \neq 0},
					\label{eq:grid_int_equal_part_int_S_not_0}
				\end{align}
				which is transitioning the sum over all cells to the sum over all particles (restricted to where $S_j \neq 0$).
				
				\begin{figure}
					\includegraphics[width=\columnwidth]{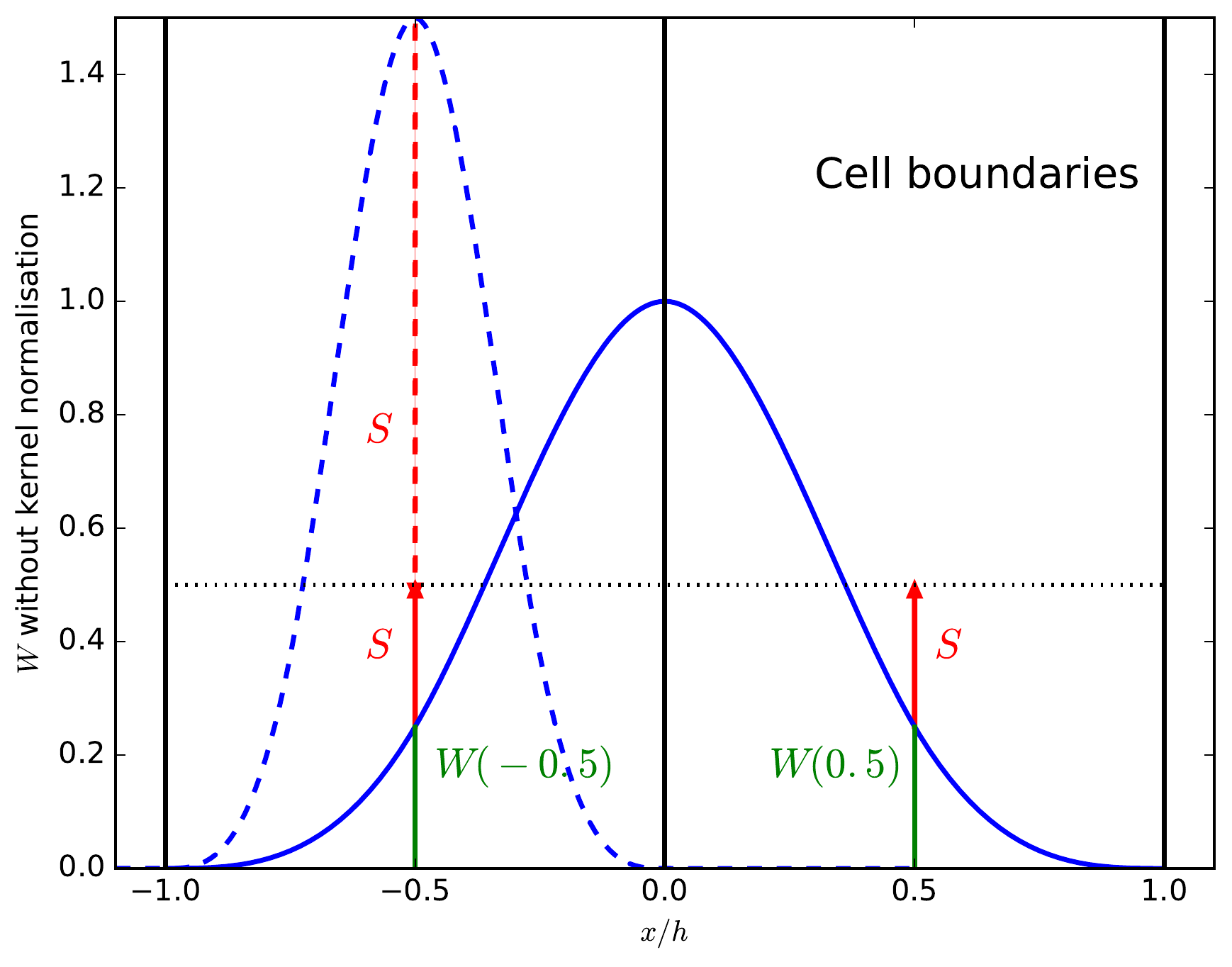}
					\caption{A simple example with one particle sitting exactly between two cells (solid blue line indicating its kernel) and one sitting exactly in a cell (dashed blue line) in order to illustrate the meaning of the S-normalisation in eq.~\ref{eq:grid_value_S_not_0}.}
					\label{fig:normSchematic}
				\end{figure}
				
				To illustrate the meaning of $S$ we sketched a dimensional example in Fig.~\ref{fig:normSchematic}. Let a particle sit at the boundary between two cells and overlap only with those two. A traditional SPH binning assigns only small portions of the particle to both cells, so that part of its information is lost ($W(-0.5) + W(0.5) \approx 0.5 \ll 1$). Since eq.~\ref{eq:diskrete_kernel_int} denotes the overlap of a particle with the grid cells, applying this correction increases the impact value of the particle to the desired amount. Furthermore, if a particle sits close to the centre of a cell and does not (or only marginally) contribute to any other cells, its impact is highly dependent on the central kernel value and, hence, its smoothing length due to the integral normalisation property of the SPH kernel. In that case a particle might even contribute too much to this grid cell, because its kernel function is bigger than unity in the central part, which leads to a reduction by $S > 1$ and eq.~\ref{eq:grid_value_S_not_0}.

				Since grid cells are of finite size there may be particles for which $S = 0$ as they \quotes{fall through the grid} by not having their kernel overlapping with any cell's centre $\vec r\,{}^{(\vec k)}$. We add those particles in a volume weighted way to the nearest cell $k$:
				\begin{align}
				\boxed{
					\left. \tilde A^{(\vec k)} \right|_{S_j = 0} := \sum_{\vec k\text{ nearest to }\vec r_j} \frac{\Delta V_j \, \left. A_j \right|_{S_j=0}}{\Delta V^{(\vec k)}} \label{eq:grid_value_S_is_0}
				}
				\end{align}
				For those it trivially holds again that
				\begin{align}
					\left. \sum_{\vec k} \Delta V^{(\vec k)} \, \tilde A^{(\vec k)} \right|_{S_j=0} = \sum_{\vec k\text{ nearest to }\vec r_j} \Delta V_j \, \left. A_j \right|_{S_j=0},
				\end{align} resulting in overall integral conservation.
	
				Furthermore, there are particles which only partially overlap with the grid, hence, eq.~\ref{eq:slimes} does not hold any more. Since such a particle's sphere of influence is cut into a peace contributing to the grid and a residual, we can not judge properly upon conserving the contribution of this particle. In this case we therefore set $S_j:=1$, i.e.\ we simply evaluate kernels at cell centres. To guarantee numerical stability we also treat small values (for example $S < 10^{-4}$) the same way as particles with $S = 0$. A different approach would be to extend the grid in order to properly calculate $S_j$. In order to be on the completely save side, this extension would have to cover the largest kernel support radius of all boundary particles. We argue that one typically extends the grid anyway, keeping all interesting features rather central while typically only a few voxels close to the boundaries are actually affected, which renders this safety procedure obsolete. About 5 voxels per side have to be added as an extension, because as soon as the smoothing lengths are much larger than the voxel sizes the discrete integral $S$ is close to one. We see this behaviour later in our test analysis section \ref{sec:conservation}. We do not concern ourselves much more with this issue here, since one can always extend the grid's size in any case where this might be desirable, without adding to much computational cost.
				
				The presented normalisation seems somewhat similar to the unity condition by \cite{Price2007}, however one has to keep in mind that the former is calculated per particle while the latter is per grid cell. This unity condition states a fundamental property of SPH (SPH noise) and does not contradict with the integral conservation condition (eq.~\ref{eq:grid_int_equal_part_int_S_not_0}) enforced by the S-normalisation, as one can easily see by calculation. It may seem that the idea of conservation and this unity condition contradict each other while they are actually compatible, however, using the SNSB method for the numerator as well as the denominator. We present the impact of this additional condition briefly in section \ref{sec:powerspectra}.

			\subsubsection{3D vs. 2D}
				\label{sec:2d}
			
				Another problem is, that the grid's resolution is strongly restricted by the availability of main memory. Assume a typical use case where we want to generate a $500^2$ sized map with a line-of-sight that requires 2000 pixels along. This results already in $4GB$ of data just for the three dimensional grid. There are possibilities like out-of-core computing to approach this problem, for example the memory management library Rambrain \citep{Imgrund2017} which is actually used by the tool SPHMapper presented in section \ref{sec:sphmapper}. However, this only allows to the user to create a huge grid and somehow fit it into memory without reducing the resulting requirements of CPU time. We propose a different approach, simplifying the problem algorithmically. Assuming one is not actually interested in the full 3D information but only in 2D data, integrating the 3D grid would be the task at hand anyway. It is actually possible to generate 2D data on the fly in order to minimise the memory footprint as well as runtime. For this approach we embed the projection task already into the formalism by using the projected kernel:
				
				\begin{align}
					\left. W \right|_{\mathrm{2D},i}(\vec x, h) &:= \int\limits_{-\infty}^{+\infty} \! \mathrm d r_i \, W\big(\vec r, h\big) \label{eq:proj_kernel} \stackrel{(\ref{eq:kernel_def})}= H^{-\nu+1} \, \left. w \right|_{\mathrm{2D}}(|\vec x|/H)
				\end{align}
				where
				\begin{align}
					\left. w \right|_{\mathrm{2D}}(b) &:= \int\limits_{-\sqrt{1 - b^2}}^{\sqrt{1 - b^2}} \! \mathrm d \ell \, w \left(\sqrt{b^2 + \ell^2}\right) \label{eq:kernel_proj_H_free}
				\end{align}
				with the impact parameter $b$. The integral $\left. w \right|_\text{2D}$ can be pre-calculated with high precision and tabulated as a function of $b$. We compare this 2D approach with the standard 3D plus integration approach in section \ref{sec:comparison} and show that it can even lead to better results than the classical 3D method due to the more precise integral along the l.o.s..
	
			\subsubsection{Non-Cartesian Grids}
			
				A big advantage of the SNSB method, and most SPH based approaches in general, is that it is totally independent of the grid structure. The only property of the grid itself which enters is the volume of each cell, therefore a regular grid is not required for the approach to work out and complexity rises only when the kernel integral over a voxel's volume is not computable in a straightforward way. Therefore, it is possible to get even better results with techniques like adaptive mesh refinement \citep{Berger1984,Berger1989} or perhaps Voronoi grids \citep{LejeuneDirichlet1850,Voronoi1908}. A typical type of binning which is commonly required is a radial or cylindrical one. While usually computed using adapted window functions, a native SPH binning would be very helpful. However, since the grid cells are not of constant size any more, typical calculations break down and constructing a method to fulfil the stated requirements is very difficult and beyond the scope of this paper.
		
	\section{Comparison by Examples}
		\label{sec:comparison}
		
		\begin{table*}
			\centering
			\begin{tabular}{ l || l | l }
				Scheme & Pros & Cons \\
				\hline \hline
				NGP & easy, fast & SPH particles treated as point-like \\
				SPH & weighting via particle properties & misses small particles, not integral conserving \\
				Broadened SPH & uses all particles & wrong weighting of particles, not integral conserving \\
				SNSB & proper weighting of all particles & computationally slightly more expensive
			\end{tabular}
			\caption{The major pros and cons of the different schemes which we compare in this section. These cover the Nearest Grid Point method and SPH binning without and with the S-norm as well as broadened SPH. If applicable, we also compare our direct 2D method to the general 3D grid followed by integration along the line of sight.}
			\label{tab:scheme_pro_con}
		\end{table*}
		
		In this section we compare some of the methods we discussed in section \ref{sec:differentmethods} by analysing a cosmological zoom simulation of a Milky Way-like galaxy at redshift zero. The methods of our choice are listed in table \ref{tab:scheme_pro_con}. We cut out a cubic box with sidelength of 600 kpc around the main galaxy and run the analysis tool Pygad (see section \ref{sec:pygad}) with different settings over the SPH data in order to generate projected maps. In order to check the errors depending on grid resolution we remove the effects of boundary particles by considering only particles with their smoothing kernel being totally enclosed by the selected box. In the following subsections we focus on the different issues we discussed in section \ref{sec:differentmethods} and illuminate how the chosen methods behave in such a common application.
		
		\subsection{The dataset}
			\label{sec:dataset}
			
			\begin{figure*}
				\centering
				\includegraphics[width=0.9\textwidth]{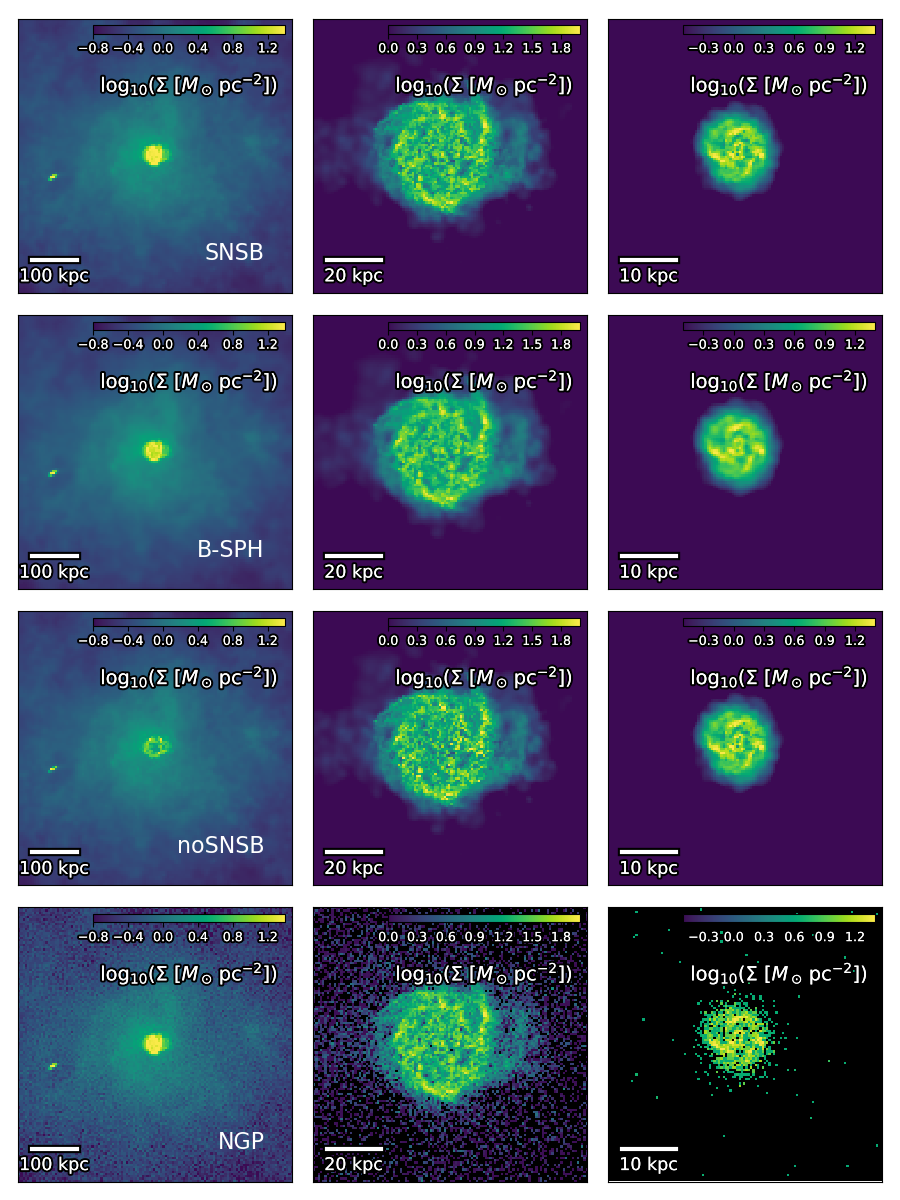}
                \caption{Projected 3D gas density maps generated with the different models for three chosen volumes. All grids have $120^3 = 1.7\times10^6$ voxels and were integrated along the same axis. The physical sizes of the boxes are $(600\,\mathrm{kpc})^3$, $(100\,\mathrm{kpc})^3$ and $(50\,\mathrm{kpc})^3$. The second column is a zoom onto the central galaxy of the first column; the last column zooms onto some satellite galaxy. The rows from top to bottom use the S-normed SPH, broadened SPH, classical SPH binning, and nearest grid point method.}
				\label{fig:mapCompMethods}
			\end{figure*}
			
            At first we show a qualitative comparison in order to understand what the data at hand contains. In Fig.~\ref{fig:mapCompMethods} we plot the column density of the chosen sample for all four methods with an underlying grid of size $120^3$. To ensure robustness of these results we carry out the same analysis also with higher (200 cells) and lower resolution (84 cells) with similar outcome and therefore do not present the additional plots. The left column shows the overall box, while the second and third one present zooms of interesting regions: One of the main galaxy itself and one of a small satellite galaxy. These three cuts contain $3.1 \cdot 10^5$, $4.1 \cdot 10^4$ and $1.4 \cdot 10^3$ particles respectively.
			
			Starting with the left column we see that while SNSB and broadened SPH agree very well, the other two methods produce quite different results. The NGP result shows correct density amplitudes but by construction lacks smoothing which leads to a rather grainy image. As outlined before, SPH data is just not interpreted correctly here. On the other hand, the classical SPH binning has a smoothed character but fails to produce proper densities in the central galaxy. Since smoothing lengths are small there compared to the voxel size, many particles just fall through the grid without being taken into account. This is quantitatively analysed later in Fig.~\ref{fig:S_val_dist}. This effect is drastically reduced in the zoom onto the central or satellite galaxy proving this to be a resolution effect. Here, the classical SPH binning and SNSB produces a fairly similar result, while the broadened SPH appears to be over-smoothed. It requires a more quantitative analysis to see properly that substructures are more prominently displayed with the SNSB method as outlined in the next subsection.
			
			The smaller the voxels, meaning the bigger the grid resolution is in comparison to the SPH resolution, the more empty pixels are produced by the NGP method and we transition from a grainy picture to an object which appears to live in a totally gas free environment. However, considering the large smoothing lengths of particles outside of the structures we know that this circumgalactic medium is not empty but just of low density. While it might have been possible to live with NGP as an approximation for the whole box, we get a totally wrong result here by misinterpreting the SPH data.
		
		\subsection{Conservation properties}
			\label{sec:conservation}
			
\begin{figure*}
				\includegraphics[width=0.99\columnwidth]{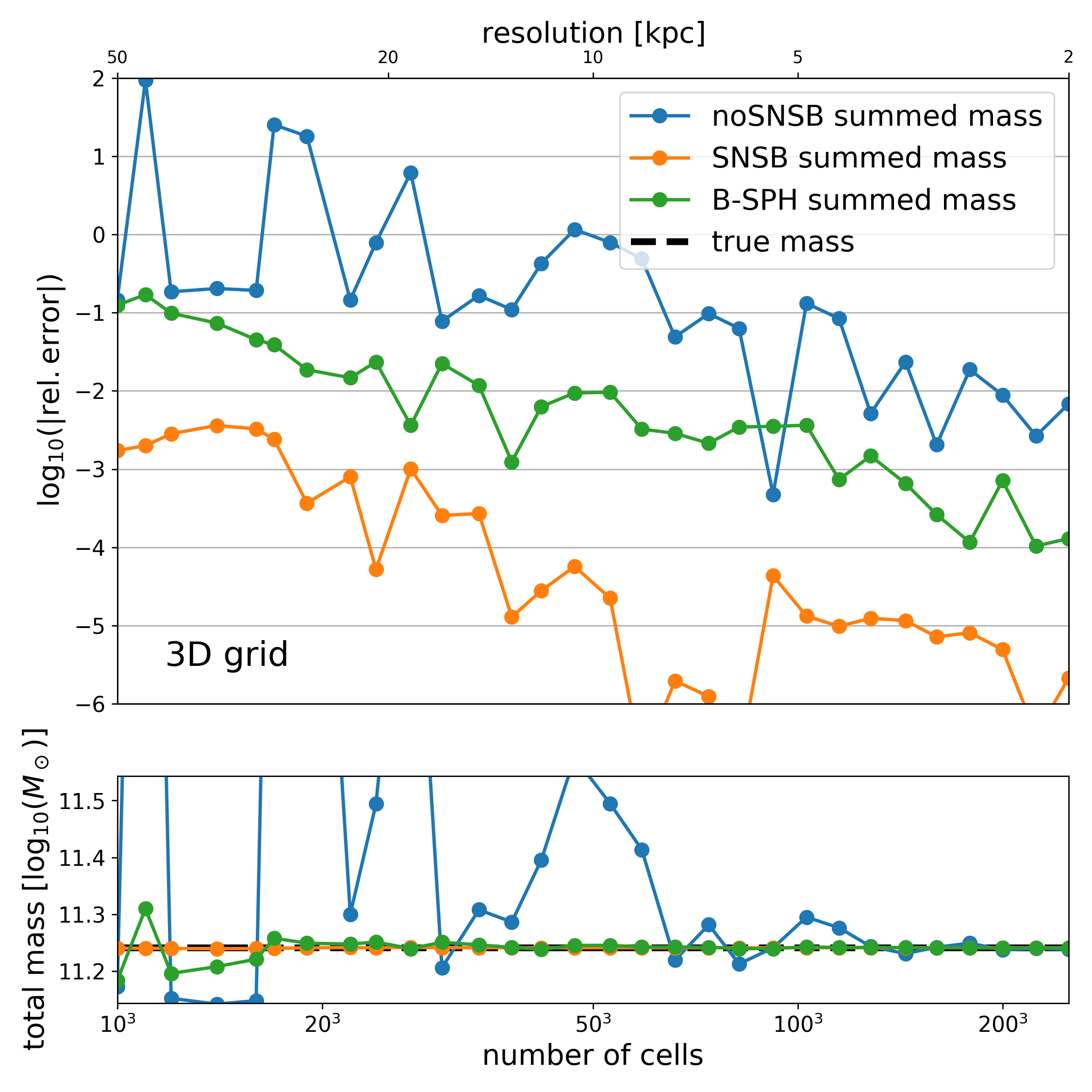}
				\includegraphics[width=0.99\columnwidth]{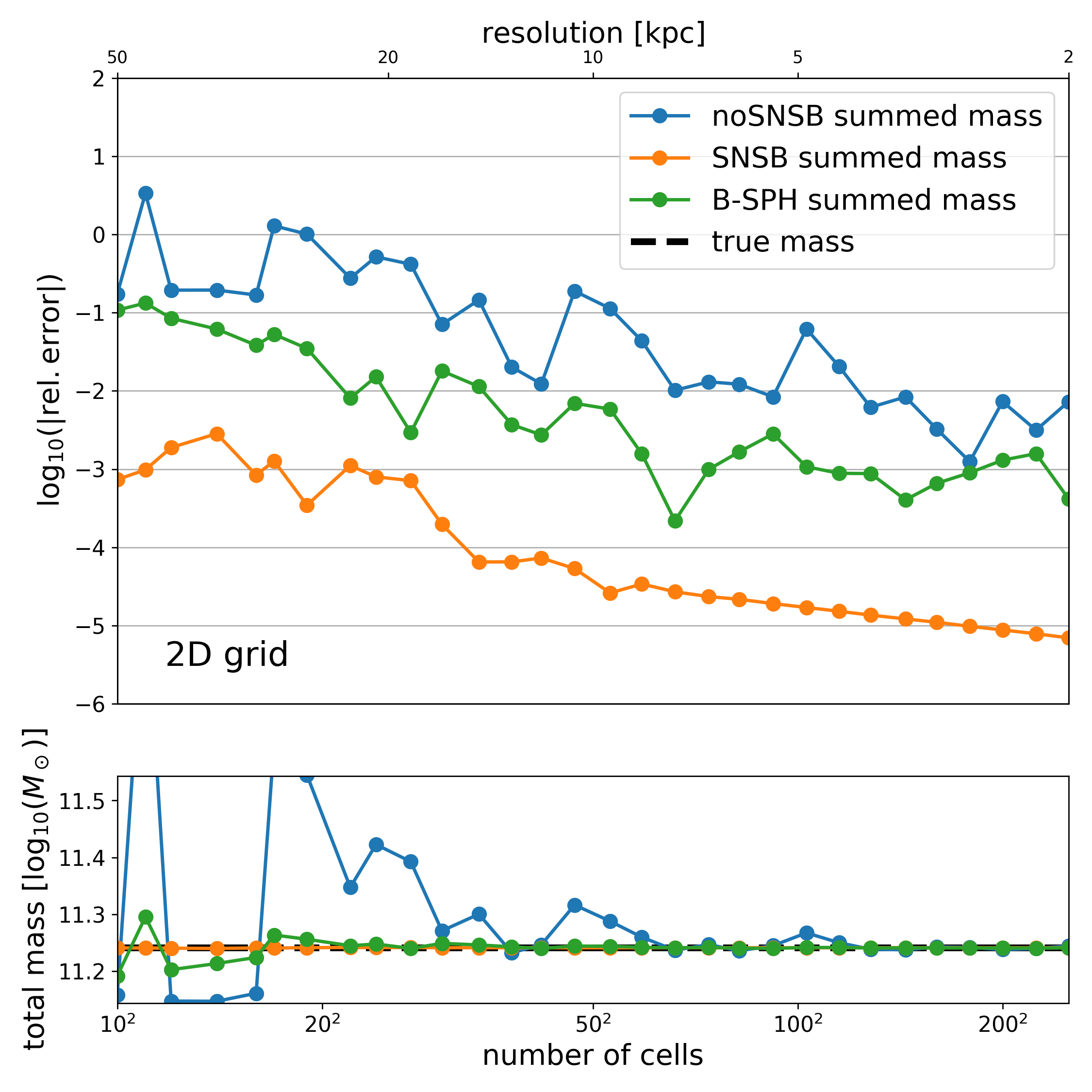}
				\caption{This figure shows the relative errors of integrated mass over the resulting grid in comparison to a sum over all particles. We vary the grid resolution from 10 to 250 cells per side. We compare our new method with and without the correction normalisation of equation \ref{eq:diskrete_kernel_int} and broadened SPH binning. The left panel displays the result for the 3D grid data while the right panel shows the projected method.}
				\label{fig:masscomp}
			\end{figure*}

			If we want to use the resulting grid data to do further calculations on it, we have to make sure that for example the total mass inside the simulation is still the same after the transformation. For that we use the full $600 \, kpc^3$ box and compare the sum over all SPH particles inside the selected region with the integral over the grid. The result is shown in Fig.~\ref{fig:masscomp} with the three dimensional methods in the left column and the two dimensional counterparts on the right. The upper row shows the relative error between the two mass calculations while the lower one displays the calculated masses.
			
			Before we come to the differences of the standard 3D approach to the native 2D version, let us focus on the left panel first. Comparing the standard SPH binning approach with the S-normalised one we can clearly see, that the latter performs much better most of the time. By construction, it is even usable for very small resolutions with only very few grid cells since it accounts for particles which do not actually overlap with any grid cell centre. The resulting relative errors are always in the sub-percent regime, while the standard SPH approach needs a minimum resolution of about 150 cells per side (which equals roughly $3kpc$ side length of a cell) to achieve even percentage error accuracy. Broadened SPH binning lies about half way between both on a logarithmic scale. Although the methods will converge eventually, changing the amount of grid cells moves all cell centres by a bit so that more resolution can sometimes even result in worse conservation. Our SNSB approach converges slightly smoother, only outperformed by the broadened SPH which is less affected by moving cell boundaries due to the broader kernel. We omit the result for NGP, since here it would just sum up all particle masses and therefore by construction conserve mass perfectly.
			
			As we can see in the lower panel, most errors lead to an overestimation of the mass in the classical SPH case which is a bit counter-intuitive. We stated that the main issue is that particles are not even taken into account and therefore, their mass is lost in the gridding process. However, one has to take an additional effect into account. Especially with a low grid resolution many substructures in the central galaxy may fall through the grid (a more quantitative analysis follows later this section) but, depending on the exact position of grid cell mid-points, some might be very close and therefore being taken into account with a strong weighting ($S \gg 1$). This leads to a grid cell with a very large value originating from only a very small particle volume inside this cell. In total it is very hard to predict, whether the result should over- or underestimate the integrated result.
			
			Now, we compare this to the right panel, where we used the native 2D prescription as presented in section \ref{sec:2d}. For classical SPH binning this decreases the error the most for all grid resolutions compared to the 3D approach. This can be easily explained by the reduction of one dimension. While the 3D methods have three space dimensions for the possibility of particles to fall through the grid, here we integrate out one dimension precisely and only two dimensions remain for the distance calculations. One can imagine this as calculating the distance perpendicular to the line of sight and not to several grid midpoints. Since this effect is most prominent in the classical binning, that method is affected the most by the 2D approach. Additionally, the native 2D version even increases the overall smoothness of convergence for all presented methods for the same reason. Since we replace the discretisation along the line of sight by a tabulated, more accurately computed integral we remove the weighting's dependence on this dimension and therefore reduce fluctuations whenever we change the grid resolution.
			
			Furthermore, an interesting effect shows up for the classical SPH binning (i.e.\ without the S-normalisation), that whenever the 2D version underestimates the mass, the 3D also underestimates it. However, this does not hold other way round. This further supports our reasoning regarding the over- and underestimation. When a particle falls through the grid in the projected plane, it will also do so in all three dimensions. But if it is not taken into account in three dimensions this might be due to the line of sight component and therefore it is accounted for in the 2D method.
			
			\begin{figure}
				\includegraphics[width=\columnwidth]{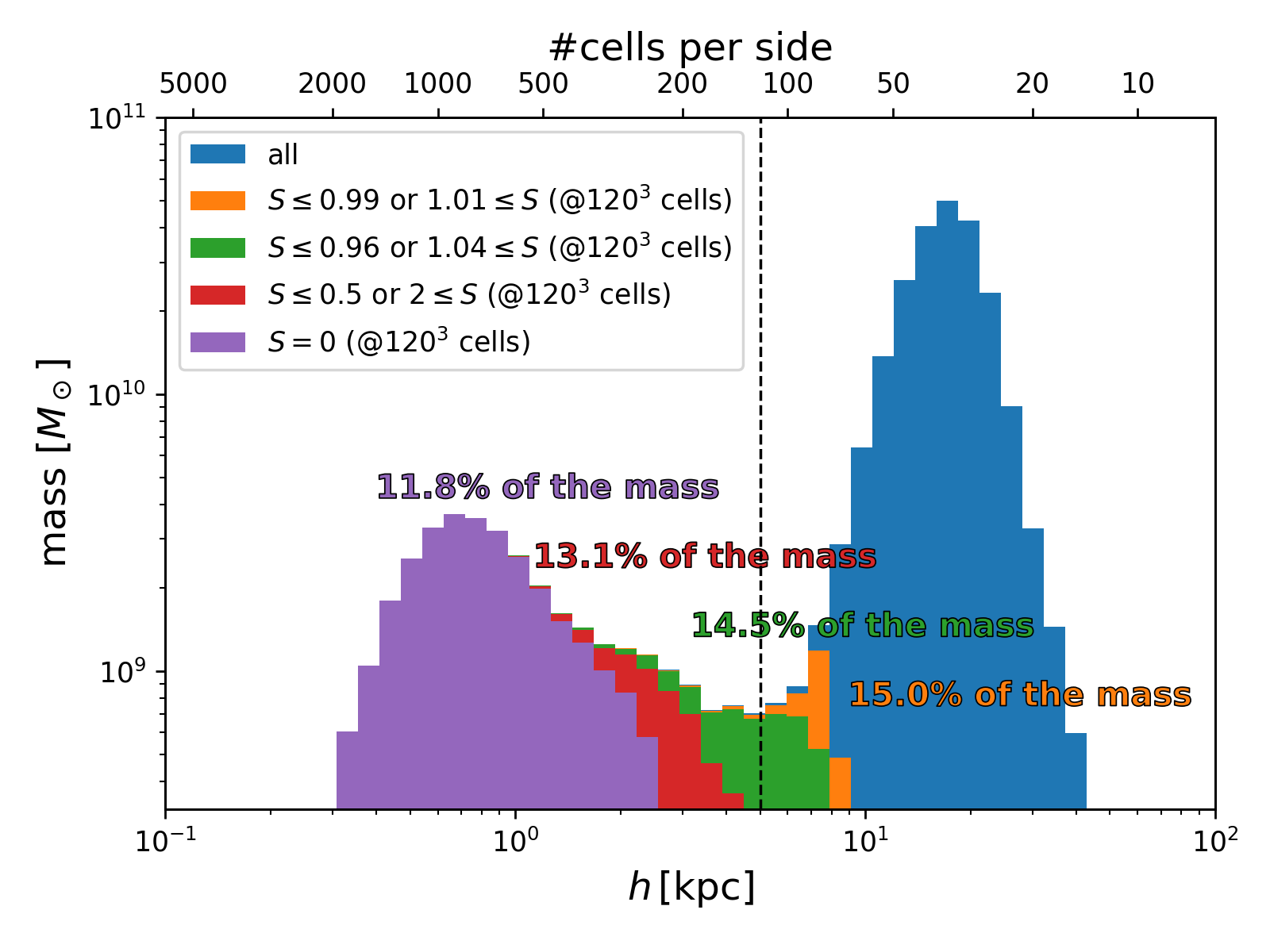}
                \caption{This histogram shows the distribution of smoothing lengths of all particles in our large box (left panels of Fig.~\ref{fig:mapCompMethods}) colour coded with different bins of $S$-values (which successively include less particles). The dashed line indicates 120 cells per side as used in Fig.~\ref{fig:mapCompMethods} and here for determining the $S$-values. Furthermore, we note down the mass contained in the respective selections for a more quantitative comparison.}
				\label{fig:hsmldist}
			\end{figure}
			
			In order to understand grid resolutions better in the context of our data set we plot the distribution of smoothing lengths in our full box binned with $120^3$ pixels in Fig.~\ref{fig:hsmldist}. To better illustrate the correspondence between grid and SPH resolution, we add the one to one relation as required number of grid cells per side on top of the plot. Since the particle masses vary slightly in our data set we plot the accumulated mass instead of counting particles. Particles in the central galaxy have smoothing lengths in the sub-$kpc$ regime and are therefore much smaller than grid cells even in our high resolution binning, implying that these are the particles which we mostly miss out in the standard SPH approach. This is especially fatal, since most of the time people are strongly interested in the high density regions. In order to fully include all particles in this simulation, an enormous grid of about $3000^3 = 2.7 \cdot 10^{10}$ cells would be required which is not feasible regarding computation time and consumed memory. Even with the native 2D variation this still results in $1.6 \cdot 10^7$ pixels. Since the grid misses out on the adaptivity of SPH this means about a factor of 100 more cells than particles in the original data set in 2D. In 3D this amounts to even $10^5$ more cells than particles. With growing SPH resolution smoothing lengths become even smaller and one would need massive computational power to solve this basic post-processing step.
			
\begin{figure}
		            	\includegraphics[width=\columnwidth]{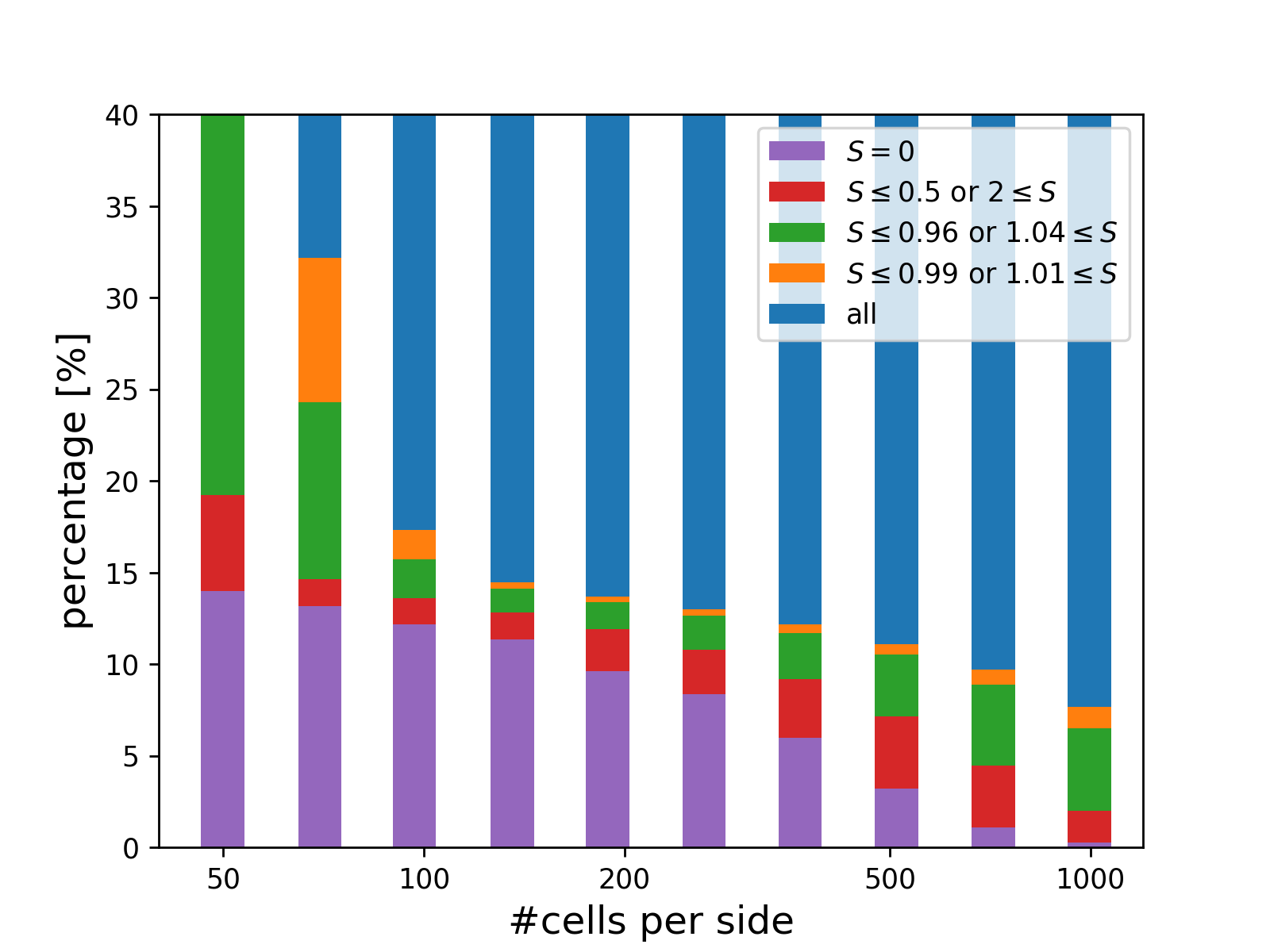} \caption{Fraction of mass with given $S$-values as a function of number of cells of the binning. (Cf.\ also Fig.~\ref{fig:hsmldist}.)} \label{fig:S_mass_fracs}
	         \end{figure}

			In Fig.~\ref{fig:S_mass_fracs} we colour code different ranges of $S$-values to better illustrate the impact of our normalisation. As denoted by the purple region about $11.8 \%$ of the mass is completely lost in a classical SPH binning, since the respective particles do not overlap with any line of sight in the given grid and therefore have $S_j = 0$. These particles are added to their respective nearest voxel according to equation \ref{eq:grid_value_S_is_0}. This exact number is of course extremely dependent on the dataset as well as on the grid resolution and position.
			
			Additionally to these particles which fall through the grid, there is also a significant component of particles of which the contribution to the grid has to be adjusted significantly. About $1.3 \%$ of mass consists of particles with $S < 0.5$ or $S > 2$ and $3.2 \%$ requires a correction of at least one percent. Of course, this leaves still $85.0 \%$ of mass almost unadjusted, but nevertheless the impact is quite significant. Furthermore, we see that particles with $S$-values deviating from 1 are the main contributors of smoothing lengths below $\approx 8 kpc$. The normalisation is therefore less necessary for particles which extend over three grid cells in each dimension. (See also the brief discussion about boundary cells earlier in section \ref{sec:problemssolutions}.)

            Fig.~\ref{fig:S_mass_fracs} displays the trend of the mass fractions within certain ranges of $S$-values in the 3D binning similar to Fig.~\ref{fig:hsmldist} but as a function of resolution. Again, it becomes clear, that lower resolution leads to higher importance of our $S$-normalisation. Grid sizes below $100^3$ are drastically affected. With this specific data set at $1000^3 = 10^9$ cells about 10\% of mass is still at least slightly affected by the normalisation. Since there are no drawbacks except for a slightly increased computational cost we can conclude that it is always worth to include our modification.
							
	         \begin{figure*}
	            	\includegraphics[width=\textwidth]{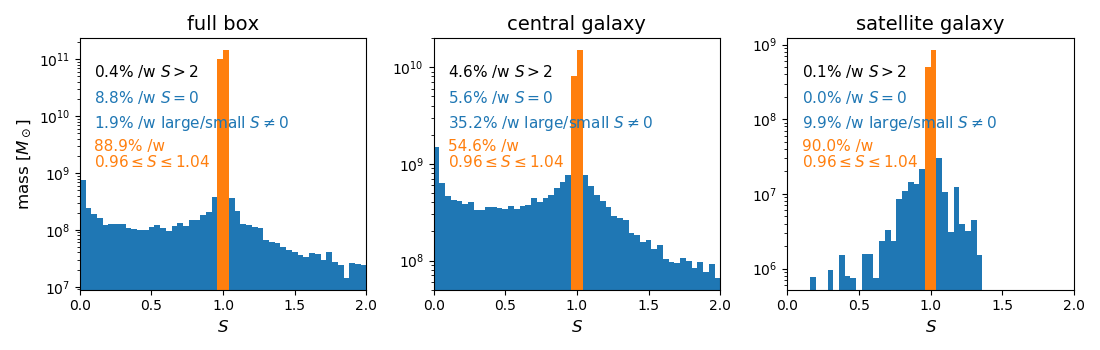}
	            	\caption{The distribution of $S$-values of the particles contributing to the grids of Fig.~\ref{fig:mapCompMethods} (first row, i.e. the fiducial method with $120^3$ cells).}
	            	\label{fig:S_val_dist}
	         \end{figure*}
	         
	         Now, we have a closer look at the distribution of $S$-normalisation values. With the histogram in Fig.~\ref{fig:S_val_dist} we can directly see the amount of contributions we loose without the re-normalisation. We plot a fine binned $S$ against the mass sitting in each bin up to $S = 2$ and note the amount of mass even beyond that. The left plot is for the full box, the middle one for the central galaxy and the right one for the small satellite. Most particles reside in this range, the ones beyond $S=2$ are the few particles with very narrow kernel functions with a very high central value which just by chance actually intersect closely with a cell centre. These particles might as well just fall through the grid, if we divide the region slightly differently into grid cells. We see, that the distribution of $S$-values is quite smooth, increasing towards the unaffected particles ($S = 1$; note that this is only about half of the particles of the central galaxy!) from both sides and for the bigger structures also towards the particles falling through ($S = 0$). We have already shown that, as expected, particles with $S = 0$ have very small smoothing lengths (see Fig.~\ref{fig:hsmldist}) in comparison to the grid size, which explains why the distribution is much narrower for the satellite galaxy, which is much smaller and therefore the effective resolution is bigger since we keep the number of grid cells fixed. The smoothing lengths of the particles in the satellite range from one to three times the cell size at this resolution.
			
		\subsection{Power-spectra}
			\label{sec:powerspectra}
		
            \begin{figure*}
                \includegraphics[width=\textwidth]{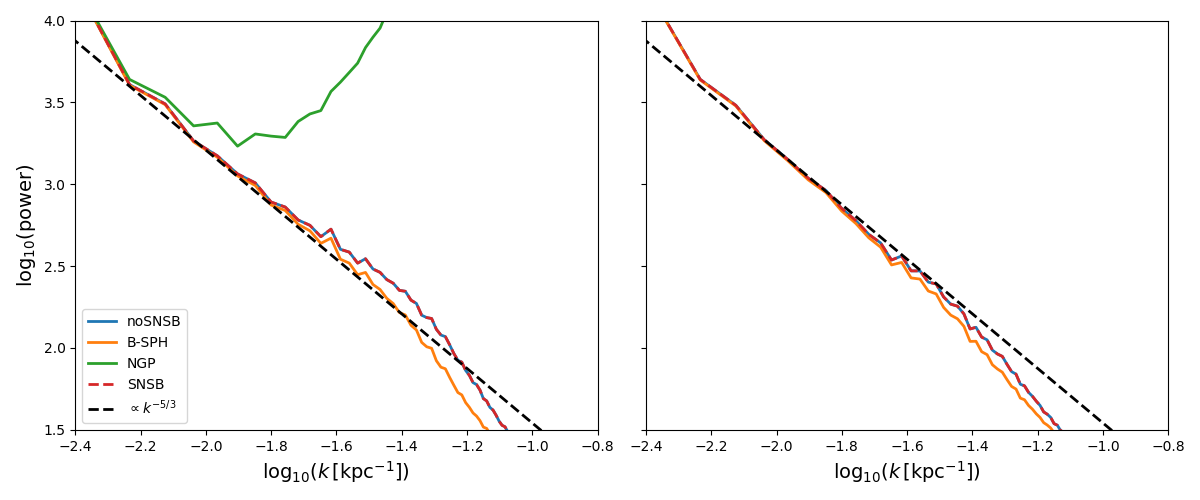}
                \caption{Kinetic power spectrum calculated from a $200^3$ sized grid for our selected methods (coloured solid lines) including the theoretical Kolmogorow spectrum with a slope of -5/3 (black dashed line).The left panel shows spectra from binning as we have presented it until now. The right panel includes an additional normalisation such that a constant field SPH also results in the same constant field on the grid (see section \ref{sec:problemssolutions} and \cite{Price2007}).}
                \label{fig:power_spec}
            \end{figure*}
            
            Finally we use our produced grids for a typical application and calculate the kinetic power spectra for the different methods. The result is shown in Fig.~\ref{fig:power_spec}. The difference between the left and the right panel is, that on the right we additionally include the unity condition eq. \ref{eq:unityconditions} as presented by \cite{Price2007}. Since the NGP method is not compatible with that, we can only include it in the left plot. We over-plot the theoretical Kolmogrov spectrum with a slope of -5/3 to have some indicator on the quality of our results, although we do not expect the simulation data to lie perfectly on top of it. First of all, we see that the NGP method only works on the very large scales (small $k$) but quickly diverges completely from the relation, showing a great amount of power on the mid range and small scales. This is the effect of the pixelation and graininess we saw in Fig.~ \ref{fig:mapCompMethods}. The classical binning and our modified version produce extremely similar results since both capture features in the structure similarly well. In contrast to them, the approach with broadened kernels is a bit off. The constant broadening plays the biggest role on small scales, where it shows up as an additional smoothing. Thus, small scale variations can vanish and the power on small scales drops. Now, in the left plot the broadened method actually produces a more convincing result but if we include the unity condition normalisation the power on small scales drops for all methods leaving the classical and $S$-normed binning closer to the expected slope.
	
	\section{Analysis packages}
		\label{sec:tools}
	
		In this final section we present two analysis packages which implement most of the discussed mapping methods. Although there exist already some publicly available tools (see for example \cite{Dolag2005,Price2007,Hummel2016}), these novel codes include not only our improved map making but also incorporate several features which make them a viable choice for fellow astrophysicists. These cover amongst others typical transformation and selection mechanisms of data, calculating of additional derived SPH quantities, combination with commonly used astrophysical codes like Cloudy \citep{Ferland2013} and in situ visualisation. Both tools cover the full range from flexibility and fast visualisation of data to the ability to process huge datasets on a regular desktop computer.

		\subsection{Pygad}
			\label{sec:pygad}

            Pygad is a Python module that provides a framework for general analysis of Gadget simulations and the basis for fast and easy development of more specific scripts and programs. It also already provides several specialised sub-modules such as one for generating mock QSO absorption lines. Pygad is publicly available at \url{https://bitbucket.org/broett/pygad} together with useful data tables and documentation.
            
            The strongest feature of Pygad is that it is designed to put the heavy lifting into the background. Loading a Gadget snapshot regardless the format is a simple call of one function with the file name as single argument. Masking the snapshot to just particles of interest is very straightforward, too, and the resulting sub-snapshot can be used like a regular one in any Pygad function. This masking can be to certain species of particles (e.g.\ gas and stars only), regions in space (e.g.\ a friend-of-friend (FoF) halo found by the built-in FoF finder or by Rockstar \citep{Behroozi2011}), or even a requirement on arbitrary properties using Python syntax (e.g.\ gas in a certain temperature range and with metallicities larger than solar).
            
            Another handy feature of Pygad we choose to highlight here are the plotting routines. They can plot any quantity, whether it is a regular block in the snapshot or a derived quantity such as stellar luminosities (provided by another module of Pygad) or certain ions (as calculated with Cloudy tables, also provided with the package). These plotting routines build on the mapping procedure presented in this work and automatically print the correct units. The units are carried with all snapshot quantities without noticeable performance penalties.
            
            As any part of Pygad, the plotting routines are kept as general and convenient as possible. And, as with any runtime critical section of Pygad, these routines are written in C++ which are called by the Python frontend in order to ensure a smooth experience. They allow to simply pass a (sub-)snapshot, a region to plot, and the name of a block, or even an Python expression of several blocks that shall be plotted.
            
            All analyses and plots in this work have been done with Pygad as it allows for quick development paired with fast runtime.
		
		\subsection{SPHMapper}
			\label{sec:sphmapper}
			
			SPHMapper is an application written in the C++11 standard and basically provides the user with a similar, although currently still more limited usability as does Pygad. With this choice of programming language comes a different approach featuring less interactivity but high applicability to huge data sets and nevertheless high versatility. As computing power grows and simulation data sets become increasingly bigger over time the main requirement for SPHMapper is to be able to handle data sets which cannot be kept in main memory all at once. To approach this issue it features the use of Rambrain \citep{Imgrund2017}, an automatic memory manager which utilises second storage capacity to overextend memory.
			
			SPHMapper is able to do all common Cartesian and radial binning tasks and is straight forward to expand further. Amongst other features, it handles snapshot reading, unit conversions, spatial transformation of the data sets, interaction with Subfind \citep{Springel2001b,Dolag2009a} outputs and selection of particles based on a complete metric in its parameter file. Binning results are visualised and output as preview using hplotlib (\url{https://github.com/sweetpony/hplotlib}).
			
			SPHMapper is still under development and will be made publicly available under an open source license soon.
		
	\section{Conclusion}
	
		In this paper we reviewed the importance of a proper understanding of SPH particle data and the difficulties of the transformation into a volume discretised picture. We showed typical approaches to solve these issues and presented a novel modification to the classical SPH binning. Our method calculates the discretised kernel integral on the grid (or a cell-particle volume ratio for particles that fall through the grid) which enables us to bin particles properties such as density in an integral conserving way (in case of density this is mass conserving).
		
		We compared our method with other classically used approaches and analysed the properties of our method using a real simulation data set of a galaxy and its satellite. We find that our method allows us to reduce the relative error in integrated mass down to $10^{-5}$ even for rather coarse grids where the classical approach is in the percent rage. We proved that our approach's main contribution comes from particles sitting in over-dense regions, where smoothing lengths are small compared to a cell's size. Highly enough resolved grids, however, are often not feasible. In our full data set, particles with a combined mass of a few percent of the total mass are significantly affected by the normalisation and over ten percent of mass is completely lost without the additional treatment of non-intersecting particles. We investigated thoroughly these contributions for different regions of our data and with different grid to particle resolution levels.
				
		Furthermore, we presented a technique to handle projection onto 2D maps already before the binning using tabulated integrals, which not only reduces the memory footprint and required computing time drastically but also reduces errors for all methods. Due to one dimension being integrated out more precisely the possibility for particles to jump between hitting a cell centre and not intersecting at all is reduced, granting much smoother convergence with number of cells.
		
		As a typical application we calculated the kinetic power spectrum of our data set and explained the behaviour of the compared techniques. As expected, divergence of the methods' results can be seen best again at small scales. We also included a quick comparison if we further include the unity condition into the analysis, which arises from the requirement that a constant field maps to a constant SPH field.
		
		We further presented two analysis tools, Pygad and SPHMapper, which incorporate the different options for binning and provide many usage features beyond that. We have shown the prowess of the $S$-normalisation and can strongly suggest to include it into any binning code.
		
		Beyond what we have shown here, improvements can be definitely made by applying adaptive or non regular grids instead of the fixed, Cartesian ones.
		
	\section{Acknowledgements}
		The authors thank K. Dolag, S. Heigl, C.-Y. Hu, F. Schulze and A. Beck for very useful discussion during the creation of this work.
		
	\section*{Bibliography}
		\bibliographystyle{apalike}
		\bibliography{sphgrid}
	
\end{document}